\documentclass[aps,prd,twocolumn,superscriptaddress,showpacs]{revtex4-1}
\usepackage{graphicx}
\pdfoutput=1
\usepackage{natbib}
\usepackage{color}
\usepackage{amsmath}
\usepackage{amssymb}
\usepackage{array}
\usepackage{multirow}
\usepackage{chemformula}
\usepackage{subcaption}

\begin{document}

\title{Nuclear and Magnetic Structures of the Frustrated Quantum Antiferromagnet Barlowite, Cu$_4$(OH)$_6$FBr}
\author{K. Tustain}
\affiliation{Department of Chemistry and Materials Innovation Factory, University of Liverpool, 51 Oxford Street, Liverpool, L7 3NY, UK}
\author{G. J. Nilsen}
\affiliation{ISIS Facility, Rutherford Appleton Laboratory, Chilton, Didcot, Oxford, OX11 0QX, UK}
\author{C. Ritter}
\affiliation{Institut Laue-Langevin, 71 Avenue des Martyrs, F-38042 Grenoble Cedex 9, France}
\author{I. da Silva}
\affiliation{ISIS Facility, Rutherford Appleton Laboratory, Chilton, Didcot, Oxford, OX11 0QX, UK}
\author{L. Clark}
\affiliation{Department of Chemistry and Materials Innovation Factory, University of Liverpool, 51 Oxford Street, Liverpool, L7 3NY, UK}
\affiliation{Department of Physics, University of Liverpool, Oliver Lodge Building, Oxford Street, Liverpool, L69 7ZE, UK}


\begin{abstract}
Barlowite, \ch{Cu4(OH)6FBr}, has attracted much attention as the parent compound of a new series of quantum spin liquid candidates, \ch{Zn$_x$Cu$_{4-x}$(OH)6FBr}. While it is known to undergo a magnetic phase transition to a long-range ordered state at $T_N=15$ K, there is still no consensus over either its nuclear or magnetic structures. Here, we use comprehensive powder neutron diffraction studies on deuterated samples of barlowite to demonstrate that the only space group consistent with the observed nuclear and magnetic diffraction at low-temperatures is the orthorhombic $Pnma$ space group. We furthermore conclude that the magnetic intensity at $T<T_N$ is correctly described by the  $Pn^\prime m^\prime a$ magnetic space group, which crucially allows the ferromagnetic component observed in previous single-crystal and powder magnetisation measurements. As such, the magnetic structure of barlowite resembles that of the related material clinoatacamite, \ch{Cu4(OH)6Cl2}, the parent compound of the well-known quantum spin liquid candidate hebertsmithite, \ch{ZnCu3(OH)6Cl2}.        
\end{abstract}

\maketitle
It is widely appreciated that the frustration, or competition, of interactions between magnetic moments in materials can result in highly unconventional magnetic behaviour \cite{Greedan2001}. The $S=1/2$ kagome antiferromagnet -- a geometrically frustrated network of equilateral triangles of antiferromagnetically coupled $S=1/2$ spins -- is a particularly notable example, as it is considered a prime candidate to host a quantum
spin liquid (QSL) ground state \cite{Mendels2016}. A QSL is an intriguing state of matter that fails to undergo classical long-range magnetic order at low temperatures \cite{Balents2010,Savary2017}. Far from being uncorrelated, QSLs are often characterized by an emergent topological gauge structure and fractional excitations, whose interesting properties make the discovery and characterisation of material realisations worthwhile and rewarding pursuits \cite{Lee2008,Tokura2017}. In particular, synthetic analogues of naturally occurring Cu$^{2+}$-based minerals \cite{Norman2016}, in which the $S=1/2$ moments of the copper ions reside on antiferromagnetically coupled kagome lattices, have provided insights into the nature of QSLs associated with strong geometric frustration. 

\begin{figure}[h]
\begin{subfigure}{1\linewidth}
\includegraphics[width=\linewidth]{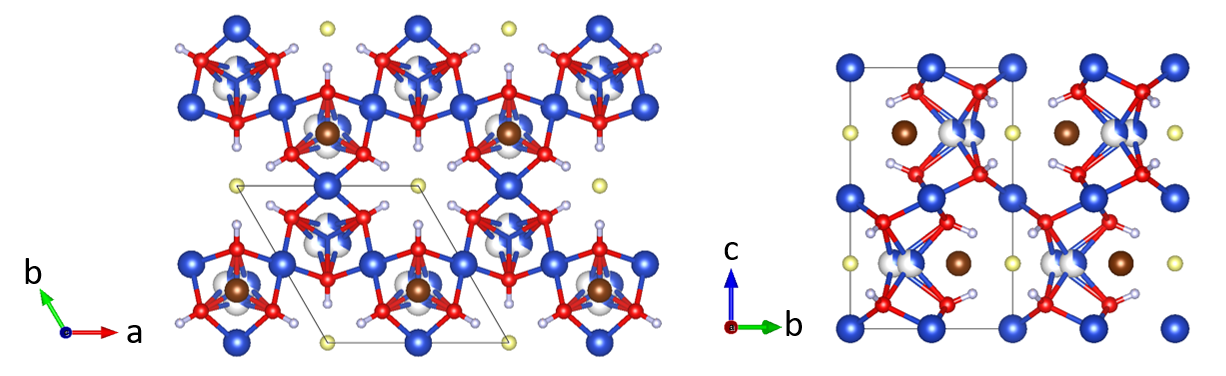}
\caption{$P6_{3}/mmc$ (300 K)}
\end{subfigure}
\begin{subfigure}{1\linewidth}
\includegraphics[width=\linewidth]{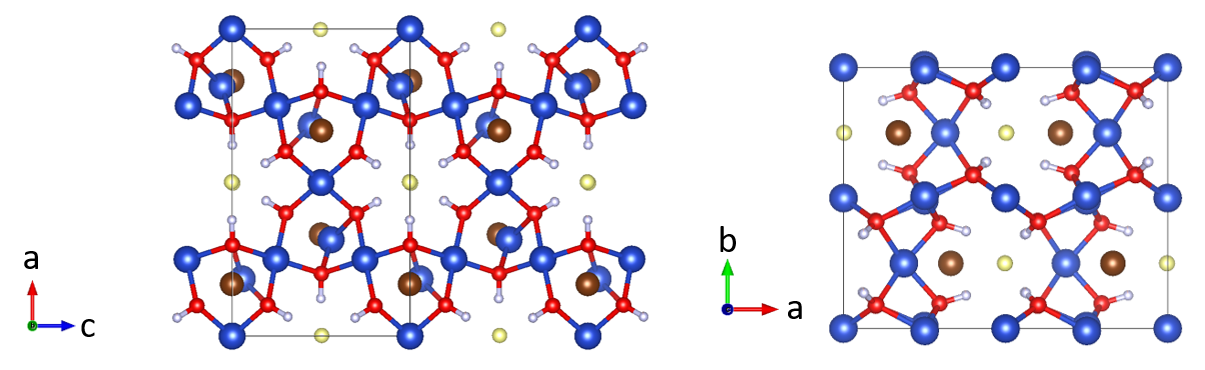}
\caption{$Pnma$ (2 K)}
\end{subfigure}\\ 
\caption{(a) At 300 K, the nuclear structure of \ch{Cu4(OD)6FBr} is described by the hexagonal $P6_{3}/mmc$ space group, where the Cu1 (blue spheres) kagome planes lie perpendicular to the $c$-axis. The Cu2 site between the kagome layers is disordered over three equivalent sites (right). (b) Below 250 K, barlowite adopts the orthorhombic $Pnma$ structure with the kagome planes now formed of two distinct copper sites in the $ac$-plane. Tilting of the intralayer Cu2 site within the kagome plane and local ordering of the interlayer Cu3 site also emerge (right). Fluorine, bromine, oxygen and deuterium atoms are shown by yellow, brown, red and white spheres, respectively.}
\label{fig:1}
\end{figure}

The most widely studied example of a Cu$^{2+}$-containing kagome mineral is herbertsmithite, \ch{ZnCu3(OH)6Cl2} \cite{Shores2005,Helton2007,Mendels2007,Bertb,DeVries2008,Freedman2010,Nilsen2013,Han2016a}. Since the exact nature of the QSL ground state in this material remains a matter of intense debate, the materials chemistry and physics communities continue to seek new candidate QSL compounds \cite{Clark2013,Balz2017,Mustonen2018}. To this end, another Cu$^{2+}$-based mineral was more recently proposed as an alternative parent compound for realising a QSL state, namely barlowite, \ch{Cu4(OH)6FBr} \cite{Elliott2014, Han2014,Jeschke2015,Han2016}. A critical issue regarding the feasibility of herbertsmithite as a model system in which to explore the QSL physics associated with the $S = 1/2$ kagome antiferromagnet model is the significant Cu$^{2+}$/Zn$^{2+}$ site disorder within its structure. This gives rise to defect spins that render the interpretation of its magnetic ground state extremely challenging \cite{Freedman2010,Nilsen2013,Han2016a}. The suggestion that the alternative stacking of the $S = 1/2$ kagome planes presented in barlowite might lead to less site disorder upon doping with Zn$^{2+}$ to favour a QSL ground state has sparked significant research efforts in synthesising and characterising barlowite and its zinc-doped derivatives \cite{Liu2015,Guterding2016,Wei2017,Feng2017,Feng,Pasco2018}. It is well documented that barlowite undergoes a magnetic ordering transition at $T_N=15$ K \cite{Han2014,Jeschke2015,Han2016,Ranjith2018,Feng,Pasco2018} but in spite of this, the nuclear and magnetic structures of barlowite are still unclear. Initial studies \cite{Han2014,Jeschke2015}, in addition to recent high-resolution synchrotron powder X-ray diffraction \cite{Smaha2018}, indicate that barlowite crystallises in a hexagonal $P6_3/mmc$ structure at room temperature. This structure, shown in Fig. {\ref{fig:1}}(a), has one copper site (Cu$1$) that forms eclipsed kagome layers in the $ab$-plane, and a second interlayer site (Cu$2$), over which the Cu$^{2+}$ ions are disordered. However, another recent report concludes that the structure of barlowite is best described by an orthorhombic $Cmcm$ model -- a subgroup of $P6_3/mmc$ -- on the basis of single-crystal X-ray and electron diffraction data collected at $100$ K and $300$ K, respectively \cite{Pasco2018}. In contrast, a powder neutron diffraction study by Feng {\textit{et al.}} indicates that the structure of barlowite below $270$ K is orthorhombic $Pnma$ \cite{Feng}. As such, the situation is unclear, and a definitive understanding of both the nuclear and magnetic structures of barlowite are required to rationalise the magnetic phase diagram of \ch{Zn$_x$Cu$_{4-x}$(OH)6FBr}. Here, we seek to address this outstanding problem through a powder neutron diffraction investigation of barlowite.

Our polycrystalline samples of deuterated barlowite, \ch{Cu4(OD)6FBr}, were sythesised via a hydrothermal reaction: \ch{CuCO3${\cdot}$Cu(OH)2} (4 mmol), \ch{CuBr2} (8 mmol) and \ch{NH4F} (12 mmol) were combined with 20 mL \ch{D2O} and sealed in a 50 mL Teflon lined stainless steel autoclave. The reaction mixture was heated at a rate of 10$^\circ$C per minute to 200$^\circ$C and held for 72 hours. The autoclave was then allowed to cool to room temperature at 5$^\circ$C per minute. The resulting turquoise product was filtered and washed several times with \ch{D2O} and dried in air. Magnetic susceptibility data were measured on a Quantum Design Magnetic Properties Measurement System (MPMS) with a SQUID magnetometer in an applied field of 1 T between $2-300$ K. Heat capacity data were measured on $2.2$ mg of pressed powder on a Quantum Design Physical Properties Measurement System (PPMS) in zero field between $2-300$ K. Time-of-flight powder neutron diffraction (PND) data were collected on the General Materials (GEM) diffractometer at the ISIS Facility of the Rutherford Appleton Laboratory \cite{GEMDOI}. Data were collected on a $1$ g sample at regular temperature intervals between $2-300$ K. Further constant wavelength PND data were recorded for a $4$ g sample on the high-intensity diffractometer D20 at the Institute Laue-Langevin, Grenoble, with ${\lambda}=2.4188$ \AA{} and monochromator take-off angle $42^\circ$ at $1.5$ K and $20$ K \cite{ILLDOI}. Nuclear and magnetic structure refinements were completed using the GSAS \cite{GSAS,EXPGUI} and FullProf \cite{FULLPROF} packages, respectively.

\begin{figure}[h]
\begin{subfigure}{1\linewidth}
\includegraphics[width=\linewidth]{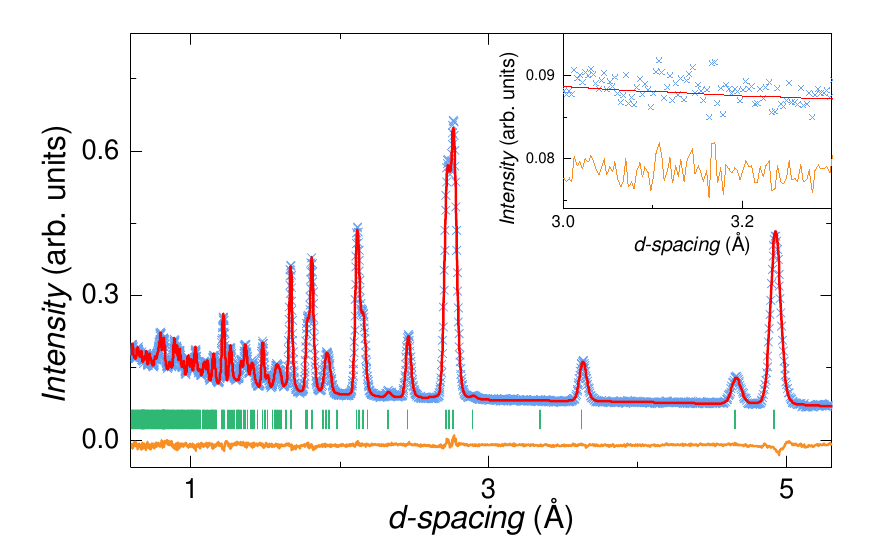}
\caption{$P6_{3}/mmc$ (300 K)}
\end{subfigure}
\hspace{8mm}
\begin{subfigure}{1\linewidth}
\includegraphics[width=\linewidth]{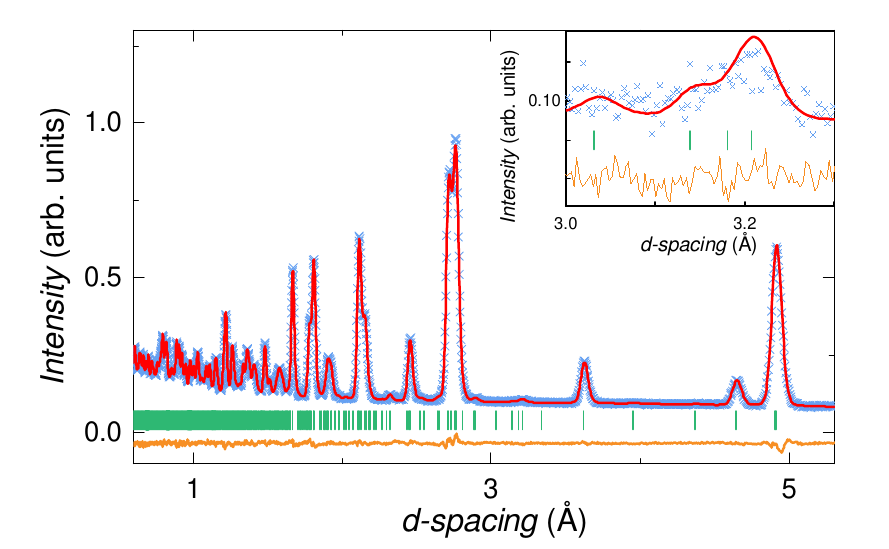}
\caption{$Pnma$ (2 K)}
\end{subfigure}\\ 
\caption{GEM Bank $3$ neutron diffraction data and Rietveld refinements of the (a) $P6_{3}/mmc$ and (b) $Pnma$ models at 300 K and 2 K, respectively. Data points are shown in blue, the fitted curves are shown in red and Bragg peak positions are represented by green tick marks. The insets show the $d$-spacing region 3.0 \AA{} - 3.3 \AA{}, where a $(102)$ reflection is observed in the $Pnma$ structure, along with additional weak $(112)$, $(311)$ and $(221)$ reflections.}
\label{fig:2}
\end{figure}

Fig. \ref{fig:2} shows the Rietveld plots of the PND data collected on Bank 3 of the GEM diffractometer at 300 K and 2 K. Our PND data collected at $300$~K are clearly well described by the hexagonal $P6_3/mmc$ model shown in Table {\ref{table:1}}, consistent with several previous studies. The Rietveld refinement of this model yields $R_{wp}=1.29$ \% and ${\chi}^2=1.83$. We find that the atomic thermal parameters within the hexagonal model are best refined anisotropically, particularly at the Cu2 site, which we attribute to the positional disorder within this model. Correspondingly, an additional proton site can be included in the refinement yielding a deuterium site occupancy of $95.9(1)$ \%. As such, this occupancy was fixed for subsequent low-temperature refinements, and reflects the high level of deuteration afforded by our synthesis procedure.

On cooling below $250$ K, we observe the appearance of additional diffraction intensity in our data, particularly in the region $3.0-3.2$ \AA{} in Bank $3$ of GEM (see Fig. {\ref{fig:2}}, insets), which is indicative of a structural phase transition. If we associate this transition with ordering of the interlayer Cu$^{2+}$ ions, it implies loss of the threefold axis and hence an orthorhombic space group. To begin exploring the possible models to describe our low-temperature data, we identified all of the orthorhombic subgroups of the two reported high-temperature structures, namely $P6_3/mmc$ and $Cmcm$, using the Bilbao Crystallographic database \cite{BCS1,BCS2,BCS3}. The full list of possible orthorhombic models is given in Table S1 in the Supplemental Material, along with the $R_{wp}$ and ${\chi}^2$ parameters obtained from Le Bail fits of each of these models to our $2$ K data. Interestingly, we identified four possible models that can reproduce the observed additional diffraction intensity in our low-temperature data; $Pbcm$, $Pmmn$, $Pmma$ and $Pnma$. All other models, including $Cmcm$, do not describe the additional low-temperature reflections, as shown in Fig. S1 in the Supplemental Material. Comparing the results of the Le Bail analyses of these four models shows that the $Pnma$ symmetry gives marginally the best overall fit. As such, we performed a Rietveld refinement of the $Pnma$ model shown in Table {\ref{table:2}} to our $2$ K data, giving a total $R_{wp} = 2.11$ \% and ${\chi}^2=2.95$. We therefore support the previous assignment of $Pnma$ below $250$ K, although, at this stage, we cannot completely exclude a lower symmetry. The fitted structure is shown in Fig. {\ref{fig:1}}(b). Compared with the $P6_3/mmc$ structure observed above $250$ K, the kagome layers now lie in the $ac$-plane and are formed of two copper sites (Cu1 and Cu2) with the latter distorted in and out of the kagome plane. This results in three different Cu-Cu bond lengths within the planes. Meanwhile, the interlayer site (now Cu3) is no longer disordered, and instead takes up a well-defined position shifted away from the Cu1 site in the $a$-direction. Presumably, it is this distortion of the kagome layers, and consequent modification of the magnetic exchanges, that allow for the magnetic ordering transition observed in barlowite below $T_N=15$ K. A summary of our refinement results across the full temperature range of our GEM experiment are given in Fig. S2 of the Supplemental Material. Thus, having now established the nuclear structures of barlowite, we next turn to the nature of the magnetically ordered state below $T_N = 15$ K. 

\begin{table*}
\caption{\label{label}Rietveld refinement nuclear structure parameters for the $P6_{3}/mmc$ model fitted to powder neutron diffraction data collected at 300 K. Refined lattice parameters are $a=b=6.6831(8)$ \AA{} and $c=9.302(1)$ \AA{} ($R_{wp}$ = 1.92 \%, $\chi^{2}=1.83)$.}.  
\begin{ruledtabular}
\hspace{-0.75cm}
\begin{tabular}{@{}rcccccccccc}
~&~&~&~&~&\multicolumn{6}{c}{$U_{aniso}$ (\AA$^{2}$)}\\
Atom & Site & $x$ & $y$ & $z$ & $U_{11}$ & $U_{12}$ &$U_{13}$ & $U_{22}$ & $U_{23}$ & $U_{33}$\\
\hline \vspace{-0.2cm} \\
Cu1 & 6g & 0.5 & 0 & 0 & 0.00724(13) & 0.00370(8) & $-$0.00228(8) & 0.00739(17) & $-$0.00456(16) & 0.01701(21)\\
Cu2 & 6h & 0.62946(8) & 0.25893(16) & 0.25 & 0.0087(4) & 0.00748(31) & 0.0 & 0.0150(6) & 0.0 & 0.0024(5) \\
F & 2b & 0 & 0 & 0.75 & 0.01511(30) & 0.00755(15) & 0.0 & 0.01511(30) & 0.0 & 0.0287(6) \\
Br & 2c & 0.6667 & 0.3333 & 0.75 & 0.01707(26) & 0.00854(13) & 0.0 & 0.01707(26) & 0.0 & 0.0089(4) \\
O1 & 12k & 0.20160(4) & 0.79840(4) & 0.90816(4) & 0.00713(12) & 0.00270(15) & $-$0.00055(8) & 0.00713(12) & 0.00055(8) & 0.01127(18)\\
D1 & 12k & 0.12430(4) & 0.87570(4) & 0.86618(4) & 0.01916(21) & 0.01321(22) & $-$0.00239(9) & 0.01916(21) & 0.00239(9) & 0.02635(26)
\label{table:1}

\end{tabular}

\end{ruledtabular}
\end{table*}

\begin{table}
\caption{\label{label}Rietveld refinement nuclear structure parameters for the $Pnma$ model fitted to powder neutron diffraction data collected at 2 K. Refined lattice parameters are $a=11.551(1$) \AA{}, $b=9.280(1)$ \AA{} and $c=6.6791(8)$ \AA{} ($R_{wp}$ = 2.11 \%, $\chi^{2}=2.949$).}  
\begin{ruledtabular}
\hspace{-0.75cm}
\begin{tabular}{@{}rccccc}

Atom & Site & $x$ & $y$ & $z$ & $U_{iso}$ (\AA$^{2}$)\\
\hline \vspace{-0.2cm} \\
Cu1 & 4a & 0 & 0 & 0 & 0.0038(5)\\
Cu2 & 8d & 0.25043(17) & 0.50911(13) & 0.24664(22) & 0.00324(25)\\
Cu3 & 4c & 0.18626(25) & 0.25 & 0.05632(27) & 0.00376(21)\\
F & 4c & 0.4979(4) & 0.25 & 0.0034(5) & 0.00973(22)\\
Br & 4c & 0.33098(27) & 0.25 & 0.49849(35) & 0.00362(16)\\
O1 & 8d & 0.29645(20) & 0.09645(21) & 0.00154(32) &  0.0025(4)\\
O2 & 8d & 0.10142(22) & 0.09037(24) & 0.1998(4) & 0.0035(4)\\
O3 & 8d & 0.40072(24) & 0.58753(22) & 0.3020(4) & 0.0053(4)\\
D1 & 8d & 0.37764(21) & 0.12684(23) & 0.9988(4) & 0.0121(6)\\
D2 & 8d & 0.06298(26) & 0.13636(24) & 0.3130(4) &  0.0126(8)\\
D3 & 8d & 0.44052(22) & 0.63892(22) & 0.19124(32) & 0.0104(6)
\label{table:2}

\end{tabular}

\end{ruledtabular}
\end{table}

Magnetic susceptibility and heat capacity data for our sample of barlowite, shown in Fig. S3 in the Supplemental Material,  clearly reveal that it undergoes a magnetic phase transition at $T_N=15$ K, consistent with all other reports. However, until now, the exact nature of the magnetically ordered state has been ambiguous. Early magnetisation measurements on a single crystal of barlowite \cite{Han2016} revealed that the magnetic ordering transition corresponds to a canted antiferromagnetic order, with a ferromagnetic component of $\sim 0.1~\mu_B$ lying in the $ab$-plane of the hexagonal $P6_3/mmc$ cell, which corresponds to the $ac$-plane of the low-temperature orthorhombic cell. A more recent PND study has revealed additional magnetic intensity at both nuclear allowed and forbidden Bragg positions below $T_N$, consistent with antiferromagnetic order with a propagation vector $\mathbf{k}=(0,0,0)$ \cite{Feng}. Rietveld analysis of the full low-temperature diffraction pattern in this previous study suggested that $\Gamma_7$ (Kovalev notation) is the active irreducible representation, and that the ordered moment is entirely suppressed on the Cu$2$ site in the kagome plane of the $Pnma$ nuclear structure. The moments on the Cu$1$ and Cu$3$ sites were both found to lie along the $a$-direction of the orthorhombic cell and have magnitudes of $0.31~\mu_B$ and $0.69~\mu_B$, respectively. This yielded a strongly suppressed average ordered moment of $0.25~\mu_B$/Cu$^{2+}$, compared to the full $S=1/2$ moment $gS=1.1~\mu_B$ \cite{Han2014}. Finally, we note that no net ferromagnetic moment was predicted in the fitted structure, which is incompatible with the aforementioned magnetisation data.
\begin{figure}[t]
\begin{center}
\includegraphics[width=\linewidth]{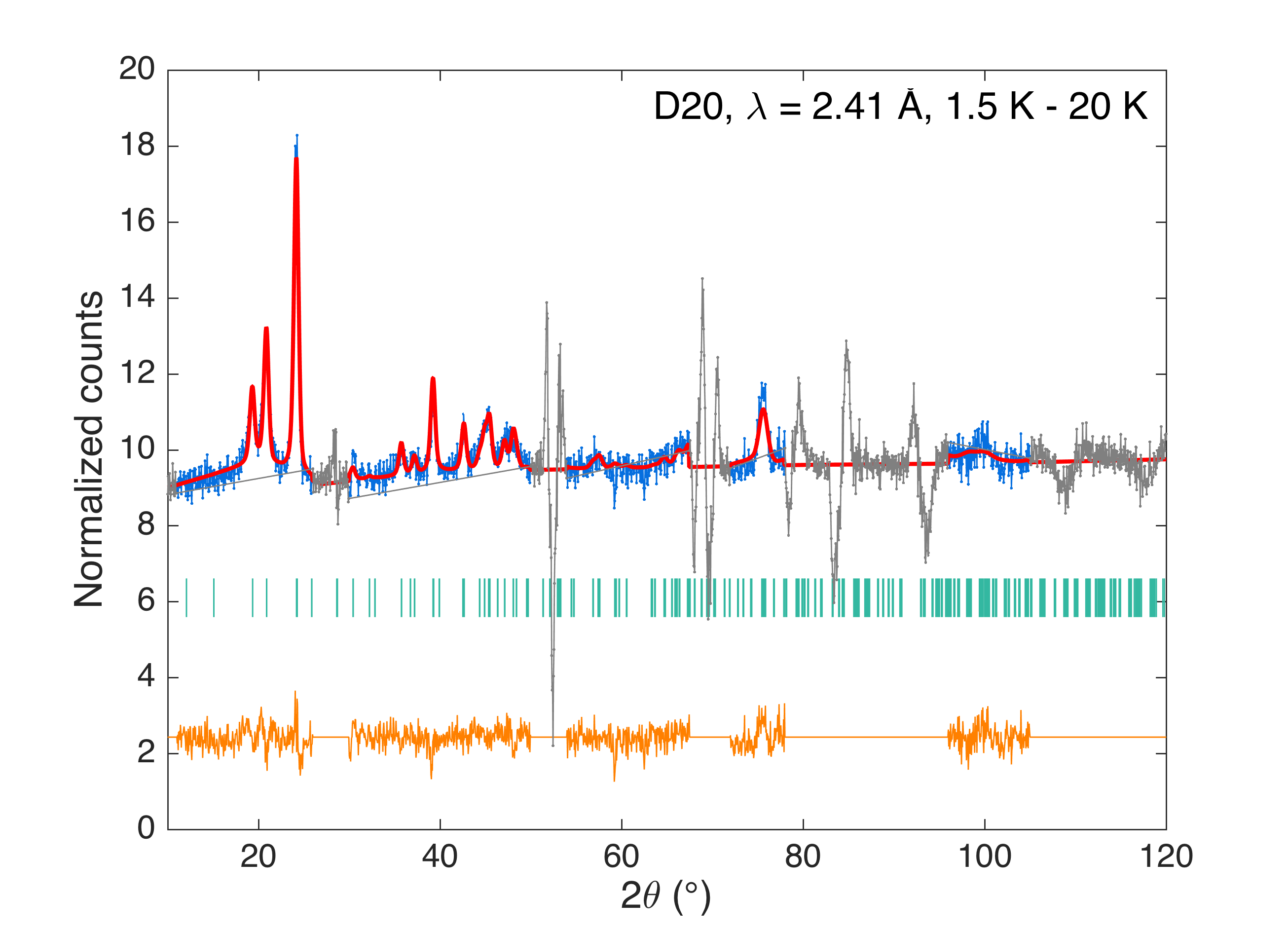}
\end{center}
\vspace{-0.4cm}
\caption{Rietveld refinement of the $Pn^\prime m^\prime a$ magnetic structure. Data points are shown in blue, the fitted curve in red, the difference in orange, and magnetic reflection positions in green. Regions of strong nuclear intensity omitted from the fit are shown in grey. Fits to a selection of other magnetic structures are shown in Fig. S4 of the Supplemental Material. The final quality factors of the refinement were $R_{wp}=2.77$ \% and $\chi^2 = 2.75$, with $11$ parameters fitted.}
\label{fig:3}
\end{figure}

\begin{figure}[h]
\begin{center}
\includegraphics[width=0.8\linewidth]{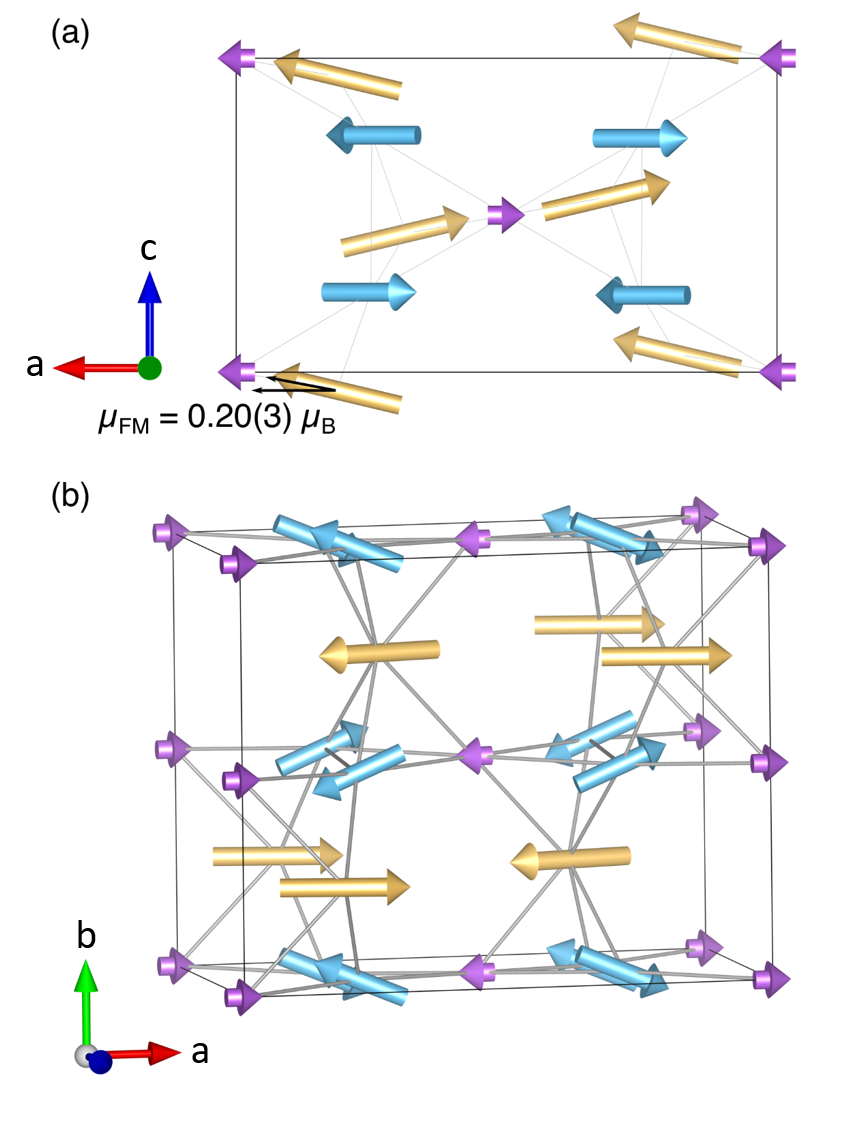}
\end{center}
\vspace{-0.4cm}
\caption{Magnetic structure of Cu$_4$(OD)$_6$FBr shown in a view along the $b$-axis (a) and in isometric view (b). The ordered moment directions on the Cu1, Cu2, and Cu3 sites are represented in purple, cyan, and orange, respectively. The ferromagnetic canting towards the $c$-axis is clearly visible in the upper panel, and corresponds to $0.20(3)~\mu_B$ per Cu3 ion, or $0.05(1)~\mu_B$ per Cu.}
\label{fig:4}
\end{figure}

Fig. {\ref{fig:3}} shows a Rietveld plot of our $T<T_N$ constant wavelength PND data collected on the D$20$ diffractometer at the ILL. Qualitatively, our data agree with those previously reported; all three previously observed magnetic peaks [$(110),~(001),~(101)$ in the orthorhombic notation] are reproduced with similar relative intensities. However, several additional magnetic peaks are observed in our data at higher angle, providing a total of ten resolved magnetic reflections for the Rietveld refinement of the magnetic structure. Furthermore, the peaks are not resolution limited -- assuming three dimensional correlations, a magnetic correlation length $\sim$ $100$ \AA{} can be estimated from the full width half-maxima. Due to the instability inherent to co-refining magnetic and nuclear structures in systems with small ordered moments, we instead use the difference of the $1.5~$K and $20$~K data for our analysis shown in Fig. \ref{fig:3}. A set of possible magnetic space groups for $\mathbf{k}=(0,0,0)$ and the $Pnma$, $Pbcn$, $Pmma$, and $Pmmn$ nuclear space groups was generated using the MAXMAGN application on the Bilbao Crystallographic Server \cite{BCS1,BCS2,BCS3,MAXMAGN}. The resulting $32$ possible magnetic space groups correspond to all possible combinations of primed and un-primed operators in each space group symbol, as can be seen for $Pnma$ in Table {\ref{table:3}}, where the corresponding irreducible representations are also given. The $24$ magnetic space groups stemming from $Pbcn$, $Pmma$, and $Pmmn$ could all be eliminated on account of poor fitting to the data. Of the remaining eight $Pnma$ magnetic space groups, only three, $Pn^\prime m a^\prime$, $Pn m^\prime a^\prime$, and $Pn^\prime m^\prime a$, allow a ferromagnetic moment in the magnetic structure, and only two allow for one in the kagome $ac$-plane. Fitting all eight possible magnetic models to the data uniquely identifies one of these latter two as the best fit, namely $Pn^\prime m^\prime a$, which is shown in Fig. {\ref{fig:4}}. While this solution has the same symmetry as that proposed in \cite{Feng}, the moment sizes are rather different, with $0.15(1)~\mu_B$, $0.41(1)~\mu_B$, and $0.53(1)~\mu_B$ on the Cu$1$, Cu$2$, and Cu$3$ sites, respectively. Furthermore, while the moments are predominantly oriented along the $a$-direction, the Cu$2$ moment cants antiferromagnetically towards the $b$-axis, while the Cu$3$ moment cants ferromagnetically towards $c$, producing a magnetization of $0.05(1)~\mu_B$ when averaged over all the Cu sites. The presence of the antiferromagnetic and ferromagnetic cantings are both statistically significant; by omitting the former, $\chi^2$ increases from $2.75$ to $3.04$, while $\chi^2$ increases to $2.88$ without the latter. This is shown in Fig. S4 of the Supplemental Material. Our model thus reproduces both the plane and approximate magnitude of the ferromagnetic component observed in previous magnetisation measurements \cite{Han2016}. We believe the large differences between our magnetic structure and that presented previously are primarily due to the choice of fitting subtracted data, rather than performing a co-refinement on the full low-temperature pattern. Indeed, an inspection of the $(101)$ peak intensity in \cite{Feng} reveals a particularly poor fit.
\begin{table*}
\caption{Magnetic space groups (in BNS notation \cite{Belov1957}) and corresponding irreducible representations (in Miller-Love notation \cite{Miller1967}), along with allowed moment directions for the $\mathbf{k}=(0,0,0)$ propagation vector in the $Pnma$ space group. Ferromagnetic components are highlighted in red. Note that in the orthorhombic structure, the kagome planes lie in the $ac$-plane.}
\begin{ruledtabular}
\hspace{-0.75cm}
\begin{tabular}{@{}rccccc}
~ & ~ & \multicolumn{3}{c}{Magnetic moment directions} & ~ \\
Magnetic space group & Irrep & Cu$_1$ & Cu$_2$ & Cu$_3$ & $\chi^2$\\
\hline \vspace{-0.2cm} \\
$Pn^\prime m^\prime a^\prime$ & $m\Gamma_1^-$ & $(0,0,0)$ & $(m_x,m_y,m_z)$ & $(m_x,0,m_z)$ & $15.2$\\
$Pn^\prime m a^\prime$ & $m\Gamma_4^+$ & $(m_x,{\color{red}m_y},m_z)$ & $(m_x,{\color{red}m_y},m_z)$ & $(0,{\color{red}m_y},0)$ & $7.3$\\
$Pn m^\prime a^\prime$ & $m\Gamma_3^+$ & $({\color{red}m_x},m_y,m_z)$ & $({\color{red}m_x},m_y,m_z)$ & $({\color{red}m_x},0,m_z)$ & $6.9$\\
$Pn^\prime m^\prime a$ & $m\Gamma_2^+$ & $(m_x,m_y,{\color{red}m_z})$ & $(m_x,m_y,{\color{red}m_z})$ & $(m_x,0,{\color{red}m_z})$ & $2.8$\\
$Pn m a^\prime$ & $m\Gamma_2^-$ & $(0,0,0)$ & $(m_x,m_y,m_z)$ & $(0,m_y,0)$ & $6.6$\\
$Pn m^\prime a$ & $m\Gamma_4^-$ & $(0,0,0)$ & $(m_x,m_y,m_z)$ & $(m_x,0,m_z)$ & $17.6$\\
$Pn^\prime m a$ & $m\Gamma_3^-$ & $(0,0,0)$ & $(m_x,m_y,m_z)$ & $(0,m_y,0)$ & $18.1$\\
$Pn m a$ & $m\Gamma_1^+$ & $(m_x,m_y,m_z)$ & $(m_x,m_y,m_z)$ & $(0,m_y,0)$ & $4.5$\\

\label{table:3}
\end{tabular}
\end{ruledtabular}
\end{table*}

Interestingly, the magnetic order in barlowite does not resemble either the 120$^\circ$ or all-in-all-out orderings expected for $\mathbf{k}=(0,0,0)$ on the kagome and pyrochlore lattices, respectively \cite{Greedan2001}. Rather, it shares several features with the closely related system clinoatacamite, Cu$_4$(OH)$_6$Cl$_2$, the parent compound of the QSL candidate herbertsmithite. It has a monoclinic $P2_1/c$ structure and, like barlowite, also contains three distinct Cu$^{2+}$ sites \cite{Wills2008}. The ordered moments in the kagome planes in clinoatacamite are nearly collinear, as in barlowite, with one of the two in-plane sites ferromagnetically aligned with the interplane site. Given the presence of six nearest-neighbour exchanges in both materials, however, it is difficult to determine whether the underlying spin Hamiltonians are similar. That being said, both magnetically ordered structures hint at a single dominant antiferromagnetic exchange in the kagome planes and ferromagnetic exchange between one in-plane site and the interplane site (here Cu1 and Cu3). In addition, the canting of the moments in both barlowite and clinoatacamite suggests that competing exchange interactions or anisotropies, either of the symmetric or Dzyaloshinskii-Moriya kind, play a key role in the formation of their magnetically ordered ground states. Indeed, in barlowite, all six possible components of the exchange tensor, and all three components of the Dzyaloshinskii-Moriya interaction are allowed for each of the six nearest-neighbour exchanges. Given the complexity of the resulting Hamiltonian, even a rough determination of its parameters must await a future single-crystal inelastic neutron scattering experiment. Should this be done for barlowite, our results suggest it could also shed new light on the long-standing problem of the Hamiltonian of clinoatacamite \cite{Zheng2005}.

In conclusion, we have presented a comprehensive powder neutron diffraction study of the nuclear and magnetic structures of the frustrated quantum antiferromagnet barlowite, \ch{Cu4(OD)6FBr}. Our analysis of time-of-flight PND data support previous proposals that the crystal structure of barlowite is hexagonal $P6_3/mmc$ and that the system undergoes a structural phase transition to an orthorhombic $Pnma$ structure on cooling below $250$ K. Crucially, however, our data do not support the notion that the crystal structure of barlowite is orthorhombic $Cmcm$ at any temperature. Moreover, we show for the first time an improved low-temperature magnetic structure of barlowite through careful analysis of subtracted, high-flux PND data, and conclude that $Pnma$ is the only space group compatible with both the nuclear and magnetic structures of this complex system. We finally note the recent report that hints at the synthesis dependence of the magnetic properties of barlowite \cite{Smaha2018}, and so sample dependence may too affect the structural properties observed by us and other groups. Nonetheless, the results presented here represent a crucial advance in our knowledge of the relationship between the nuclear and magnetic structures of this material, that will ultimately lay crucial foundations for the understanding of the QSL phase reported to emerge in related zinc-doped barlowite phases \cite{Wei2017,Feng2017}.

LC acknowledges the University of Liverpool for start-up support and a studentship to KT. Work at the ISIS Facility and the Institut Laue-Langevin was supported by the STFC. We also acknowledge the ISIS Deuteration Facility for the provision of \ch{D2O} for the synthesis of deuterated samples.
\newpage
\newpage
\bibliographystyle{apsrev4-1}
\bibliography{PhysRevMat2018}
\clearpage
\setcounter{figure}{0}
\setcounter{table}{0}
\renewcommand{\thefigure}{S\arabic{figure}}
\renewcommand{\thetable}{S\arabic{table}}
\onecolumngrid
\appendix
\section{Supplemental Material}
\begin{figure*}[h]
\centering
\includegraphics[width=0.4\linewidth]{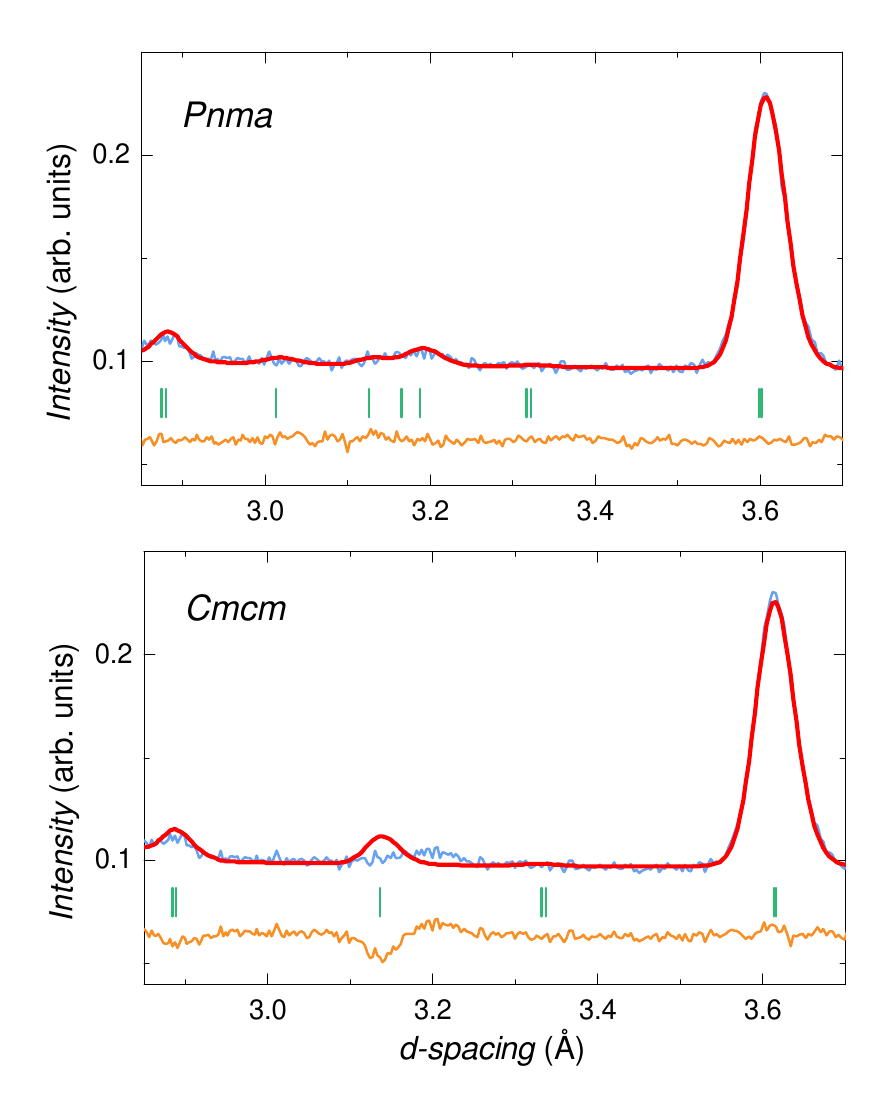}
\caption{GEM Bank 3 neutron diffraction data and Le Bail fits of the $Pnma$ (top panel) and $Cmcm$ (bottom panel) models for data collected at 2 K. $R_{wp}$ and $\chi^{2}$ values for each bank can be found in Table \ref{table:lebailtable}.}
\label{fig:S1}
\end{figure*}

\begin{figure*}[h]
\centering
\includegraphics[width=\linewidth]{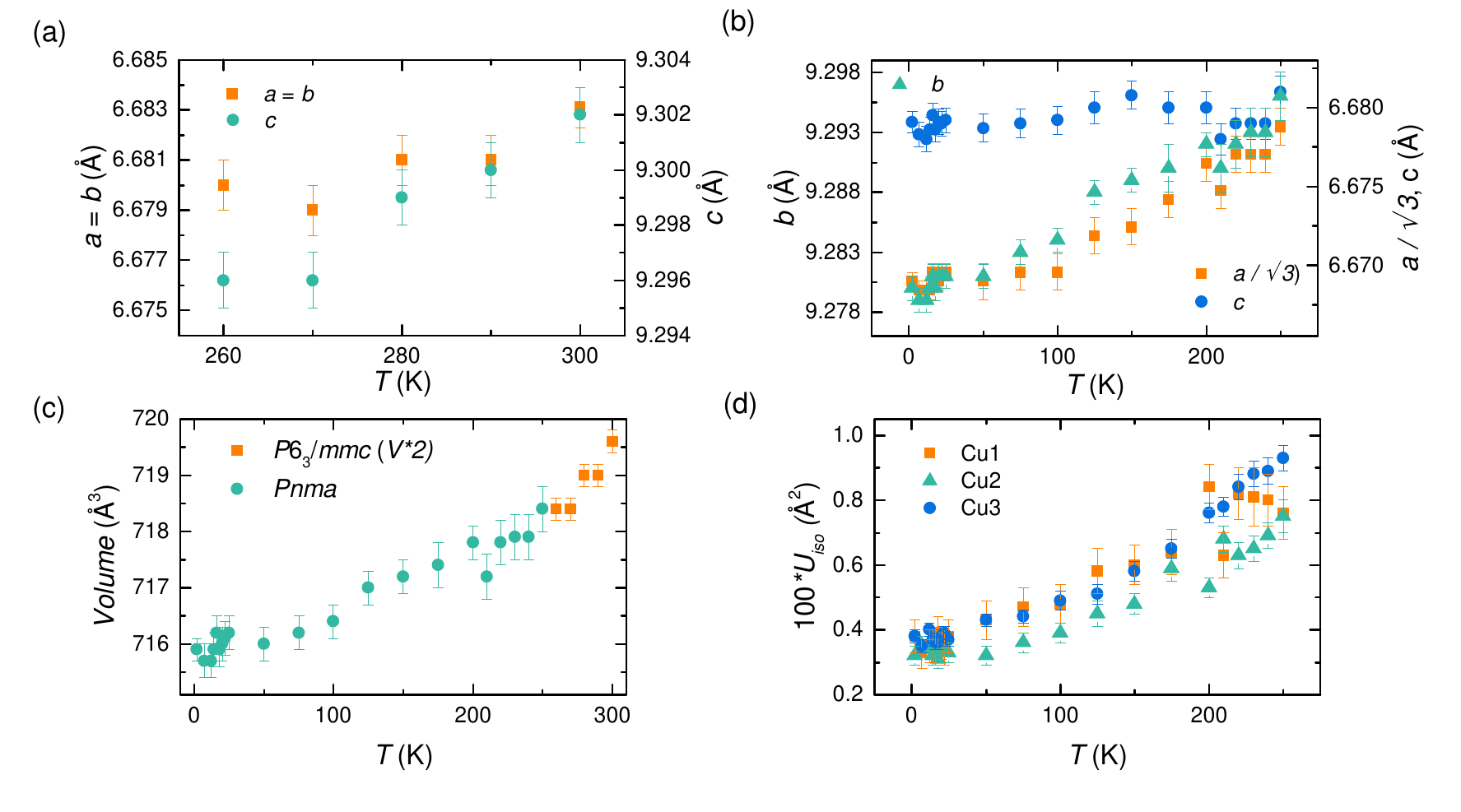}
\caption{Refined lattice parameters for (a) the $P6_{3}/mmc$ model between $260-300$ K and (b) the $Pnma$ model between $2-250$ K. The corresponding change in volume for the full temperature range is shown in (c). In (d), the isotropic thermal parameters for the three copper sites in the $Pnma$ model between $2-250$ K are shown.}
\label{fig:S2}
\end{figure*}

\begin{figure*}[h]
\centering
\includegraphics[width=0.6\linewidth]{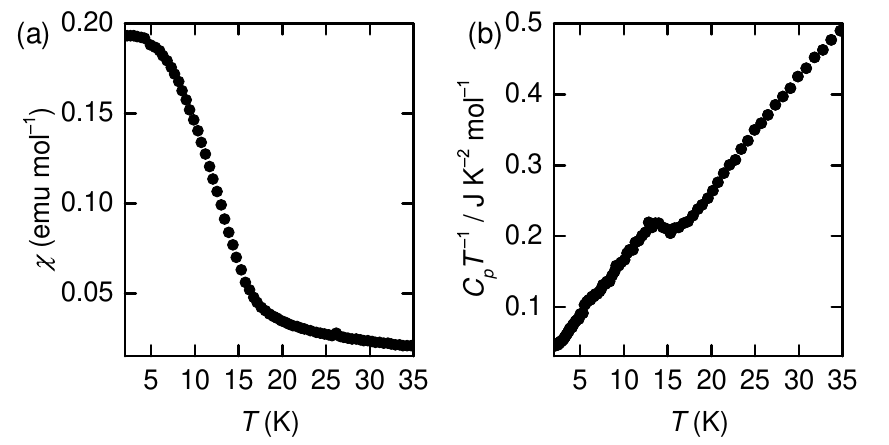}
\caption{(a) Zero-field cooled magnetic susceptibility measured in an applied field of 1 T and (b) $C_{p}T^{-1}$ measured in zero field for \ch{Cu4(OH)6FBr}. Both indicate a transition to long-range magnetic order at $T_{N} = 15$ K. Fitting a Curie-Weiss model to the magnetic susceptibility data between $150-300$ K yields a Weiss temperature $\theta=-112.3(3)$ K and a Curie constant $C=1.762(2)$ emu mol$^{-1}$ K. The latter corresponds to an effective magnetic moment $\mu_{eff}=3.75(1) \mu_{B}$ per formula unit.}
\label{fig:S3}
\end{figure*}

\begin{figure*}[h]
\begin{center}
\includegraphics[width=0.4\linewidth]{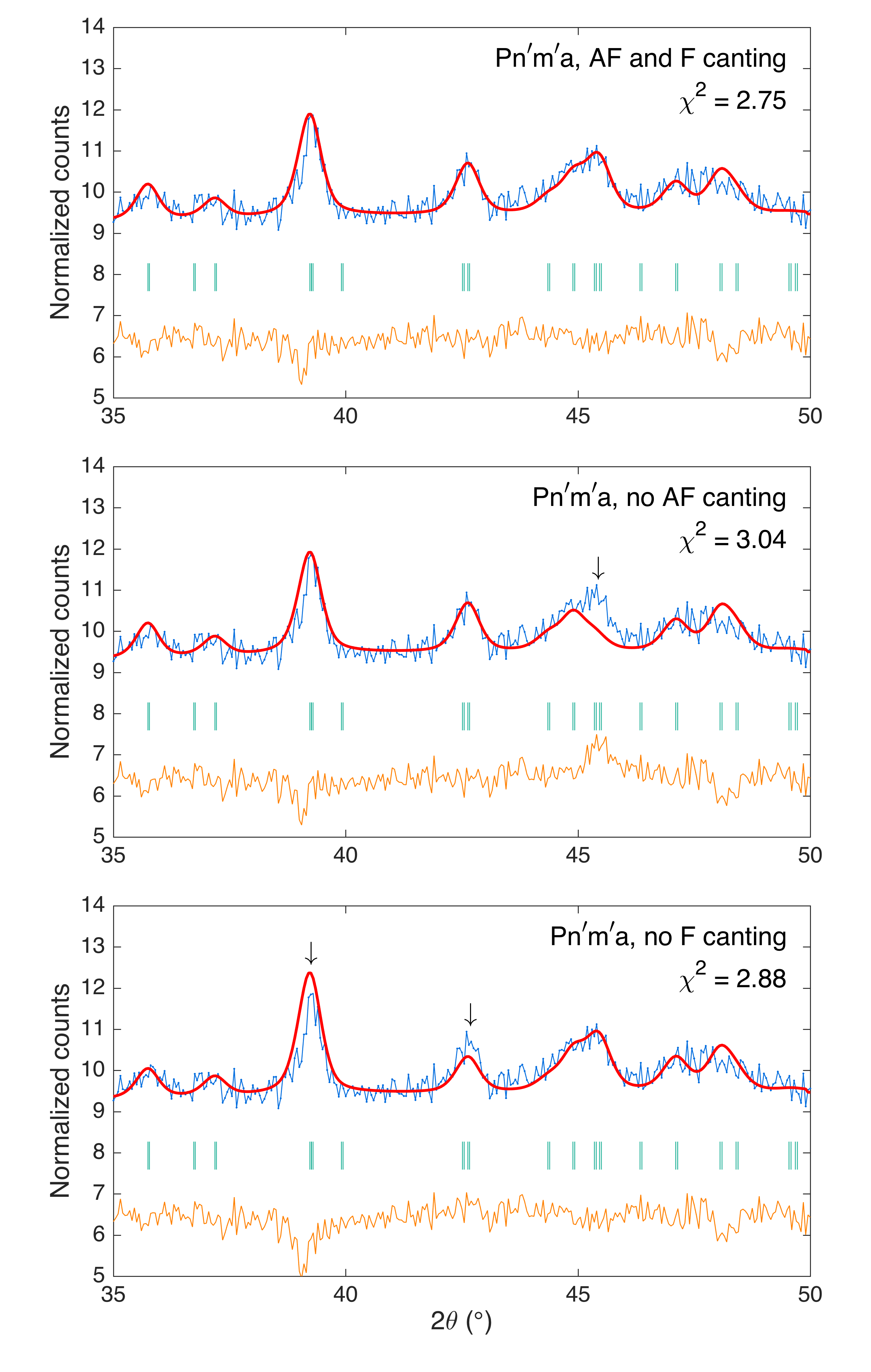}
\end{center}
\vspace{-0.4cm}
\caption{Rietveld refinements to the $Pn^\prime m^\prime a$ magnetic structure with and without the antiferromagnetic (AF) $b$-axis and ferromagnetic (F) $c$-axis canting on the Cu2 and Cu3 sites, respectively. In the lower two panels, the arrows mark reflections where the fit quality becomes poorer compared to the unconstrained model.}
\label{fig:S4}
\end{figure*}

Table \ref{table:S1} shows the $R_{wp}$ and $\chi^{2}$ values for Le Bail fits of each of the orthorhombic subgroups of $P6_{3}/mmc$ and $Cmcm$ space groups to data collected on each bank of GEM at 2 K. Fig. \ref{fig:S1} shows Le Bail fits of the $Pnma$ and $Cmcm$ models to data collected on Bank 3 of GEM at 2 K, where the former clearly accounts for the additional Bragg reflections we observe in low-temperature data. Fig. \ref{fig:S2}(a)$-$(c) show lattice parameters calculated for the $P6_{3}/mmc$ model between $260-300$ K and the $Pnma$ model between $2-250$ K, in addition to the corresponding volumes. Fig. \ref{fig:S2}(d) shows the isotropic thermal parameters determined for each Cu site in the $Pnma$ model. Magnetic susceptibility and heat capacity data measured for \ch{Cu4(OH)6FBr} are shown in Fig. \ref{fig:S3}. Both show that barlowite undergoes a transition to long-range magnetic order at $T_{N} = 15$ K. Table \ref{table:S2} gives the refined magnetic moment components for the unconstrained $Pn^\prime m^\prime a$, and Table \ref{table:S3} the transformation of the magnetic moment for all 16 Cu positions in the asymmetric unit. Fig. \ref{fig:S4} shows a comparison between the fitted model with and without antiferromagnetic and ferromagnetic cantings.

\begin{table*}[!h]
\caption{\label{label}Summary of the $R_{wp}$ and $\chi^{2}$ values obtained in each bank from Le Bail fits to powder neutron diffraction data collected for \ch{Cu4(OD)6FBr} at 2 K. }.  
\begin{ruledtabular}
\hspace{-0.75cm}
\begin{tabular}{@{}rcccccccc}
~&\multicolumn{7}{c}{$R_{wp}$(\%)}&~\\
Model & Bank 1 & Bank 2 & Bank 3 & Bank 4 & Bank 5 & Bank 6 & Total & $\chi^{2}$\\
\hline \vspace{-0.2cm} \\
$Cmcm$ & 4.03 & 2.54 & 2.27 & 2.33 & 2.30 & 1.76 & 2.27 & 2.378\\
$C222_{1}$ & 4.06 & 2.62 & 2.40 & 2.36 & 2.49 & 1.66 & 2.34 & 2.535\\
$Cmc2_{1}$ & 4.02 & 2.57 & 2.48 & 3.12 & 4.63 & 3.36 & 3.50 & 5.679\\
$Pbcm$ & 4.13 & 2.55 & 2.16 & 2.27 & 2.26 & 1.45 & 2.18 &
2.201\\
$Pbcn$ & 4.05 & 2.53 & 2.19 & 2.24 & 2.26 & 1.43 & 2.17 & 2.180\\
$Amm2$ & 4.19 & 2.57 & 2.30 & 2.27 & 2.39 & 1.53 & 2.25 & 2.341\\
$Pmmn$ & 4.01 & 2.50 & 2.00 & 2.08 & 2.15 & 1.31 & 2.04 & 1.917\\
$Ama2$ & 4.07 & 2.63 & 2.42 & 2.37 & 2.51 & 1.67 & 2.35 & 2.557\\
$Pmma$ & 4.08 & 2.47 & 2.11 & 2.15 & 2.15 & 1.31 & 2.08 & 2.002\\
$Pnna$ & 3.98 & 2.56 & 2.15 &  2.26 & 2.33 & 1.49 & 2.20 & 2.237\\
$Pnma$ & 3.97 & 2.49 & 1.98 & 2.09 & 2.15 & 1.30 & 2.03 & 1.914\\
$Pnnm$ & 4.10 & 2.49 & 2.07 & 2.08 & 2.16 & 1.32 & 2.06 &  1.954
\label{table:S1}
\end{tabular}
\end{ruledtabular}
\end{table*}

\begin{table*}[h]
\caption{The refined magnetic moment components for the unconstrained $Pn^\prime m^\prime a$ model.}
\begin{ruledtabular}
\hspace{-0.75cm}
\begin{tabular}{@{}rll}
Site & Position & Magnetic moment components \\
~ & ~ & $[m_x,m_y,m_z]$ \\
\hline \vspace{-0.2cm} \\
Cu1 & $(0,0,0)$ & $[0.144(9),0.02(2),-0.05(2)]$ \\
Cu2 & $(x,y,z)$ & $[0.361(7),0.19(1),0.405(8)]$ \\
Cu3 & $(x,1/4,z)$ & $[0.493(9),0,0.20(3)]$
\label{table:S2}
\end{tabular}
\end{ruledtabular}
\end{table*}

\begin{table*}[h]
\caption{The transformations of the magnetic moment with the symmetry operations of $Pnma$ in the magnetic space group $Pn^\prime m^\prime a$.}
\begin{ruledtabular}
\hspace{-0.75cm}
\begin{tabular}{@{}rll}
Site & Position & Mag. moment direction \\
\hline \vspace{-0.2cm} \\
Cu1 & $(0,0,0)$ & $(m_x,m_y,m_z)$ \\
$(4a)$ & $(1/2,0,12)$ & $(-m_x,-m_y,m_z)$ \\
~ & $(0,1/2,0)$ & $(m_x,-m_y,m_z)$ \\
~ & $(1/2,1/2,1/2)$ & $(-m_x,m_y,m_z)$ \\
Cu2 & $(x,y,z)$ & $(m_x,m_y,m_z)$ \\
$(8d)$ & $(-x,y+1/2,-z)$ & $(-m_x,-m_y,m_z)$ \\
~ & $(-x,y+1/2,-z)$ & $(m_x,-m_y,m_z)$ \\
~ & $(x+1/2,-y+1/2,-z+1/2)$ & $(-m_x,m_y,m_z)$ \\
~ & $(-x,-y,-z)$ & $(m_x,m_y,m_z)$ \\
~ & $(x+1/2,y,-z+1/2)$ & $(-m_x,-m_y,m_z)$ \\
~ & $(x,-y+1/2,z)$ & $(m_x,-m_y,m_z)$ \\
~ & $(-x+1/2,y+1/2,z+1/2)$ & $(-m_x,m_y,m_z)$ \\
Cu3 & $(x,1/4,z)$ & $(m_x,0,m_z)$ \\
$(4c)$ & $(-x+1/2,3/4,z+1/2)$ & $(-m_x,0,m_z)$ \\
~ & $(-x,3/4,-z)$ & $(m_x,0,m_z)$ \\
~ & $(x+1/2,1/4,-z+1/2)$ & $(-m_x,0,m_z)$ \\
\label{table:S3}
\end{tabular}
\end{ruledtabular}
\end{table*}

\end{document}